\documentstyle[sprocl,psfig]{article}

\catcode`\@=11
\def\@citexnew[#1]#2{\if@filesw\immediate\write\@auxout
        {\string\citation{#2}}\fi
\def\@citea{}\@citenew{\@for\@citeb:=#2\do
        {\@citea{}\def\@citea{,}\@ifundefined
        {b@\@citeb}{{\bf ?}\@warning
        {Citation `\@citeb' on page \thepage \space undefined}}
        {\csname b@\@citeb\endcsname}}}{#1}}

\newif\if@cghi
\def\citelow{\@cghifalse\@ifnextchar [{\@tempswatrue
        \@citexnew}{\@tempswafalse\@citexnew[]}}
\def\@citenew#1#2{{$\![#1]$\if@tempswa\typeout
        {IJCGA warning: optional citation argument 
        ignored: `#2'} \fi}}

\catcode`\@=12
\setbox0=\hbox{\hbox{-}\,}

\def\st{\scriptstyle}

\def\be{\begin{equation}}
\def\ee{\end{equation}}
\def\bea{\begin{eqnarray}}
\def\eea{\end{eqnarray}}

\def\be{\begin{equation}}
\def\ee{\end{equation}}
\def\bea{\begin{eqnarray}}
\def\eea{\end{eqnarray}}
\def\simlt{\stackrel{<}{{}_\sim}}
\def\simgt{\stackrel{>}{{}_\sim}}

\def\NPB#1#2#3{{\it Nucl.~Phys.} {\bf{B#1}} (19#2) #3}
\def\PLB#1#2#3{{\it Phys.~Lett.} {\bf{B#1}} (19#2) #3}
\def\PRD#1#2#3{{\it Phys.~Rev.} {\bf{D#1}} (19#2) #3}
\def\PRL#1#2#3{{\it Phys.~Rev.~Lett.} {\bf{#1}} (19#2) #3}

\def\MPLA#1#2#3{{\it Mod.~Phys.~Lett.} {\bf#1} (19#2) #3}

\def\AP#1#2#3{{\it Ann.~Phys.} {\bf#1} (19#2) #3}

\def\HPA#1#2#3{{\it Helv.~Phys.~Acta} {\bf#1} (19#2) #3}
\def\JETPL#1#2#3{{\it JETP~Lett.} {\bf#1} (19#2) #3}

\def\IJMPD#1#2#3{{\it Int.~J.~Mod.~Phys.} {\bf D#1} (19#2) #3}

\def\mst1{m_{\;\widetilde{t}_{1}}}

\def\st{\;\widetilde{t}}

\def\mst2{m_{\;\widetilde{t}_{2}}}
\def\mst12{m_{\;\widetilde{t}_{1,2}}}

\def\msb1{m_{\;\widetilde{b}_{1}}}
\def\msb2{m_{\;\widetilde{b}_{2}}}
\def\msb12{m_{\;\widetilde{b}_{1,2}}}

\def\mtilde2{\widetilde{m}^{2}}

\relax

\begin{document}

\begin{flushright}
IEM-FT-188/99\phantom{000000}\\
IFT-UAM/CSIC-99-10\phantom{000000}\\
hep-ph/9903274\phantom{000000}\\ 
\vspace{1eM}
March 1999\phantom{000000}
\end{flushright}

\vspace{.3cm}
\title{ELECTROWEAK BARYOGENESIS IN THE MSSM~\footnote{Plenary talk given at 
Strong and Electroweak Matter (SEW98), 3 December 1998, Copenhagen.} }

\author{M.~QUIROS, M.~SECO}

\address{Instituto de Estructura de la Materia,\\ 
E-28006, Madrid, SPAIN\\e-mail: mariano@makoki.iem.csic.es}


\maketitle\abstracts{We review the baryogenesis scenario in the MSSM at the
perturbative level and, in particular, the impact of two-loop corrections on
the strength of the phase transition and the amount of generated baryon
asymmetry. We confirm the baryogenesis window, where $m_H\simlt 115$ GeV, for
$m_Q\simlt$ a few GeV, and the right-handed stop mass is constrained in the
region, $100\, GeV\simlt m_{\widetilde{t}}\simlt m_t$. This scenario will be
tested at LEP and Tevatron colliders.}

\section{Introduction}
Electroweak baryogenesis~\cite{baryogenesis} is an appealing mechanism to 
explain the observed 
value of the baryon-to-entropy ratio, $n_B/s\sim 10^{-10}$, at the
electroweak phase transition~\cite{reviews}, that can be tested at 
present and future 
high-energy colliders. Although the Standard Model (SM) contains all the
necessary ingredients~\cite{baryogenesis} for a successful baryogenesis, 
it fails in providing enough baryon asymmetry. In particular it has been
proven by perturbative~\cite{AndH}$^{\copy0}$\cite{twoloop} and
non-perturbative~\cite{nonpert} methods that, for Higgs masses allowed by
present LEP bounds, the phase transition is too weakly first order or does
not exist at all (it is an analytical cross-over~\cite{crossover}), and any
previously generated baryon asymmetry would be washed out after the phase
transition. On the other hand the amount of CP violation arising from the CKM
phase is too small for generating the observed baryon asymmetry~\cite{CPSM}.
Therefore electroweak baryogenesis requires physics beyond the Standard Model
at the weak scale.

Among the possible extensions of the Standard Model at the weak scale, its
minimal supersymmetric extension (MSSM) is the best motivated one. It
provides a technical solution to the hierarchy problem and has deep roots 
in more fundamental theories unifying gravity with the rest of interactions.
As for
the strength of the phase transition~\cite{early}$^{\copy0}$\cite{mariano2}, a region
in the space of supersymmetric parameters has been 
found~\cite{CQW}$^{\copy0}$\cite{CQW2} where the phase transition is strong enough to let
sphaleron interactions go out of equilibrium after the phase transition and
not erase the generated baryon asymmetry. This region (the so-called 
light stop scenario) provides values of the
lightest Higgs and stop eigenstate masses which will be covered at LEP2 and
Tevatron colliders.

The MSSM has new violating phases~\cite{CPMSSM} that can drive enough
amount of baryon asymmetry~\cite{CQRVW}$^{\copy0}$\cite{last} provided that 
the previous phases are not much
less than 1 and the charginos and neutralinos are not heavier than 
200 GeV. In all calculations of the baryon asymmetry the details of the wall
parameters play a prominent role in the final result. In particular the wall
thickness, $L_\omega$, and the relative variation of the two Higgs fields
along the wall, $\Delta\beta$, are typical parameters which 
the generated baryon asymmetry depends upon. Although reasonable assumptions
about the Higgs profiles along the wall have been done, as e.g. kinks or
sinusoidal patterns interpolating between the broken and the symmetric phases,
as well as estimates on the value of $\Delta\beta$ based on purely potential
energy considerations, it is clear that the reliability of those estimates
as well as more precise computations of the baryon asymmetry should rely on
realistic calculations of the Higgs profiles and the tunneling processes from
the false to the true vacuum. Such a task, achieved in the case
of one Higgs field in the Standard Model~\cite{Brihaye93}, 
has been recently done in the case of two-Higgs doublets of the MSSM
in Refs.~\citelow{mqs,cm,pj}.

In this talk we will review the impact of two-loop corrections in the MSSM
effective potential both for the strength of the phase transition and for the
calculation of the bubble parameters. We will show that they
produce an enhancement of both the strength of the phase transition (leading
to the possibility of encompassing higher values of the Higgs mass) and of the
amount of generated baryon asymmetry (leading to alleviated bounds on the
value of the CP-violating parameters).

\section{Phase transition: two-loop enhancement}

\begin{figure}[htb]
\centerline{
\psfig{figure=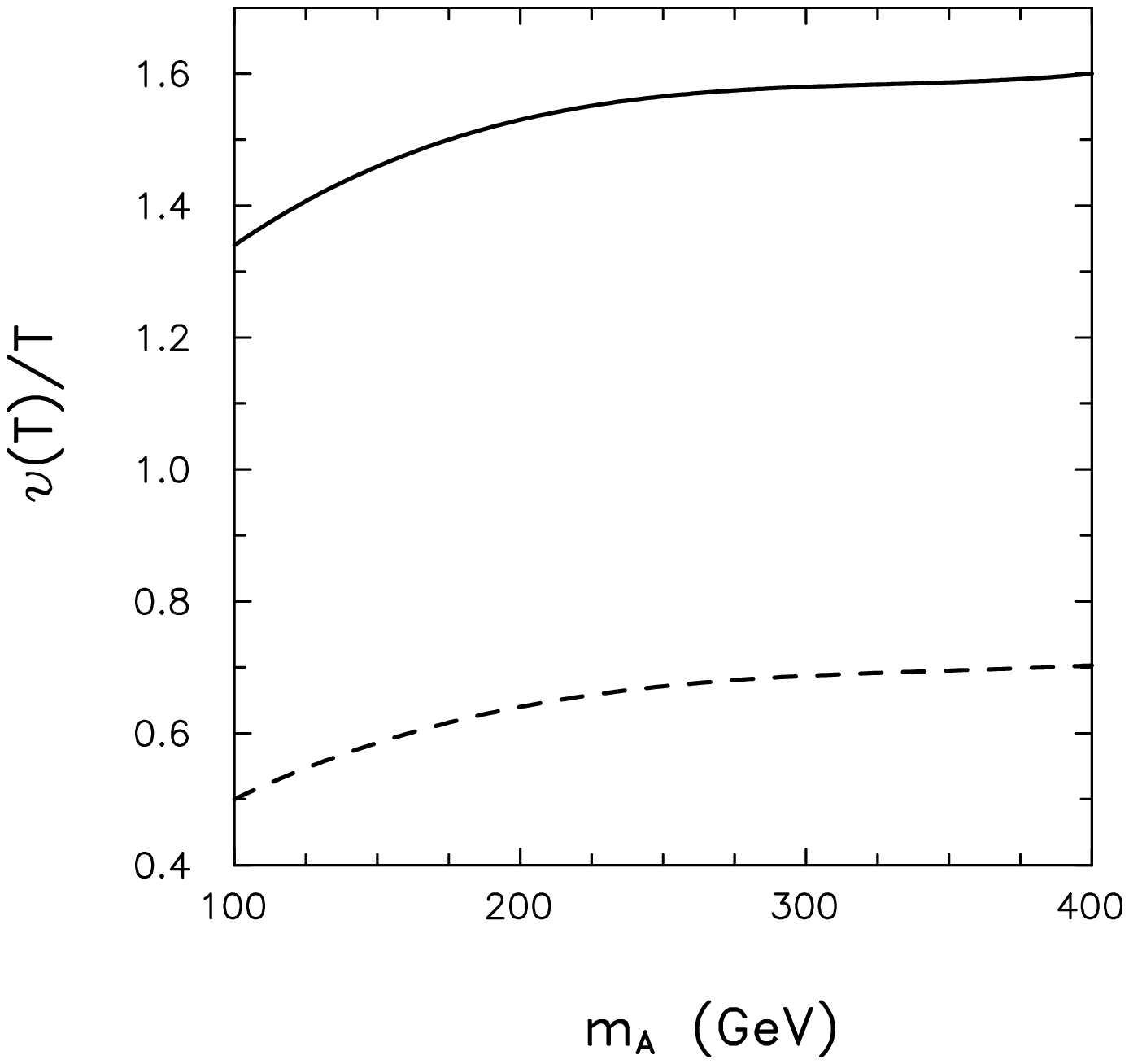,height=6cm,bbllx=4.cm,bblly=3.5cm,bburx=18.cm,bbury=16.5cm}
\psfig{figure=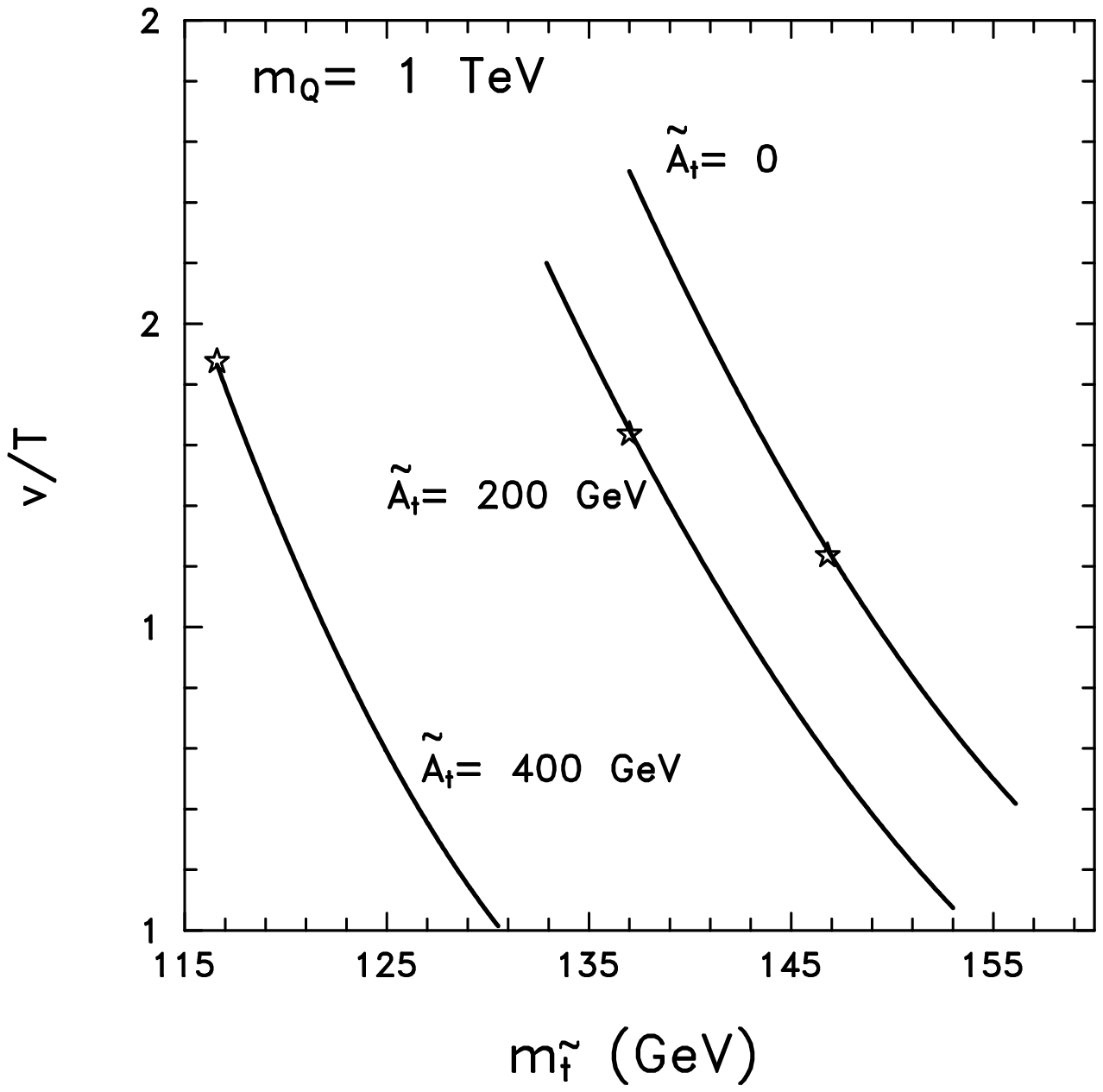,height=6.1cm,bbllx=4.5cm,bblly=3.2cm,bburx=18.5cm,bbury=16.2cm}}
\caption{Left panel: $v(T_c)/T_c$ as a function of $m_A$ in the two-loop
(solid) and one-loop (dashed) approximations for $m_Q=1$ TeV, $\tan\beta=2.5$, 
$m_{\widetilde{t}_R}=150$ GeV  and $\widetilde{A}_t=0$. Right panel:
$v(T_c)/T_c$ as a  function of $m_{\widetilde{t}_R}$ for $m_Q=1$ TeV,
$\tan\beta=3$ and various values of the mixing. The value of the Higgs mass
varies between $\sim$ 90 and 95 GeV.}
\label{fig1} 
\end{figure}

The possibility of achieving, in the MSSM, a strong-enough phase transition 
for not washing out any previously generated baryon asymmetry, characterized
by the condition
\begin{equation}
v(T_c)/T_c\simgt 1\quad ,
\label{condicion}
\end{equation}
has been recently strengthened by three facts:
\begin{itemize}
\item
The presence of light $\widetilde{t}_R$ (with small mixing $\widetilde{A}_t$
considerably enhances the strength of the phase
transition~\cite{mariano1,mariano2,CQW,Delepine}. This is the so-called light
stop scenario. 
\item
Two-loop corrections enhance the phase transition in the SM~\cite{twoloop},
and in the MSSM~\cite{JoseR,JRB,Schmidt,CQW2}.
\item
The validity of perturbation theory, and in particular the results of two-loop
calculations, for the light stop scenario has been recently confirmed by
non-perturbative results~\cite{npMSSM}.
\end{itemize}
In Fig.~\ref{fig1}, left panel, we show the two- and one-loop approximation
for the order parameter $v(T_c)/T_c$ as a function of $m_A$ for the values of
supersymmetric parameters which are indicated in the caption. We can see that
the one-loop approximation does not satisfy condition (\ref{condicion}), and
the corresponding case would be ruled out, while the two-loop
approximation does; two-loop corrections are able to
rescue this case.

However the price the light stop scenario has to pay is that it may require
moderately negative values of the supersymmetric parameter
$m_U^2\equiv-\widetilde{m}_U^2$ and then, apart from the electroweak minimum
along the Higgs ($\phi$) direction, another color breaking minimum along the
$U\equiv\widetilde{t}_R$ direction might appear. Therefore both
directions should be studied at finite temperature. A preview of
our results is shown in Fig.~\ref{fig1}, right panel, where
$v(T_c)/T_c$ is plotted versus $m_{\widetilde{t}_R}$ for
different values of the mixing parameter. We can see that, as
anticipated, $v(T_c)/T_c$ increases when the stop-right mass
decreases, for all values of $\widetilde{A}_t$. However, for 
given values of $m_{\widetilde{t}_R}$, which are indicated with a
star in the plot, the electroweak minimum stops being the true
minimum at the temperature $T_c$ and the phase transition
proceeds first toward the color breaking minimum along the $U$
direction. For instance for $\widetilde{A}_t=0$ this happens at
$m_{\widetilde{t}_R}=147$ GeV, while for $\widetilde{A}_t=400$
GeV it happens at $m_{\widetilde{t}_R}=117$ GeV. Moreover 
all lines stop at some value of $m_{\widetilde{t}_R}$ where the
electroweak minimum becomes unstable.

To systematically analyze the different possibilities we have
computed the two-loop effective potential along the $\phi$ and
$U$ directions and compared their cosmological evolutions with
$T$. The two-loop effective potential along the $\phi$ direction
was carefully studied in Refs.~\citelow{JoseR,JRB}. The one-loop
correction is dominated by the exchange of the top/stop sector
while the two-loop effective potential is given by two-loop 
diagrams with stops and gluons, as well as one-loop diagrams with
the stop thermal counterterm. The two-loop effective potential
along the $U$ direction was studied in Refs.~\citelow{Schmidt,CQW2}. 
The mass spectrum is given in table 1, where we have 
considered small mixing $\widetilde{A}_t/m_Q$ and large gluino masses.
\begin{table}[t]
\caption{Mass spectrum along the $U$ direction.
\label{spectrum}}
\vspace{0.2cm}
\begin{center}
\footnotesize
\begin{tabular}{|c|c|c|}
\hline &&\\
field & d.o.f. &  mass$^2$ \\ \hline &&\\
4 gluons &12& $g_s^2\,U^2/2$\\
1 gluon &3& $2g_s^2\, U^2/3$ \\
1 $B$ gauge boson &3& $\, g'^2 U^2/9$\\ 
5 squark-goldstones &5& $m_U^2+g_s^2\, U^2/3$ \\
1 squark &1 & $m_U^2+g_s^2\, U^2$ \\
\hline &&\\
4 $\widetilde{Q}_L$-Higgs &4& $-m_H^2/2+h_t^2\sin^2\beta U^2$\\
2 Dirac fermions ($t_L,\widetilde{H}$) &8&$\mu^2+h_t^2 U^2$\\
\hline
\end{tabular}
\end{center}
\end{table}
The one-loop diagrams correspond to the propagation of gluons,
squarks and Higgses as well as Dirac fermions. The leading
two-loop contributions correspond to sunset and figure-eight
diagrams with the fields of table 1 propagating, as well as
one-loop diagrams with thermal counterterms insertions
corresponding to gluons and squarks.

\begin{figure}[htb]
\centerline{
\psfig{figure=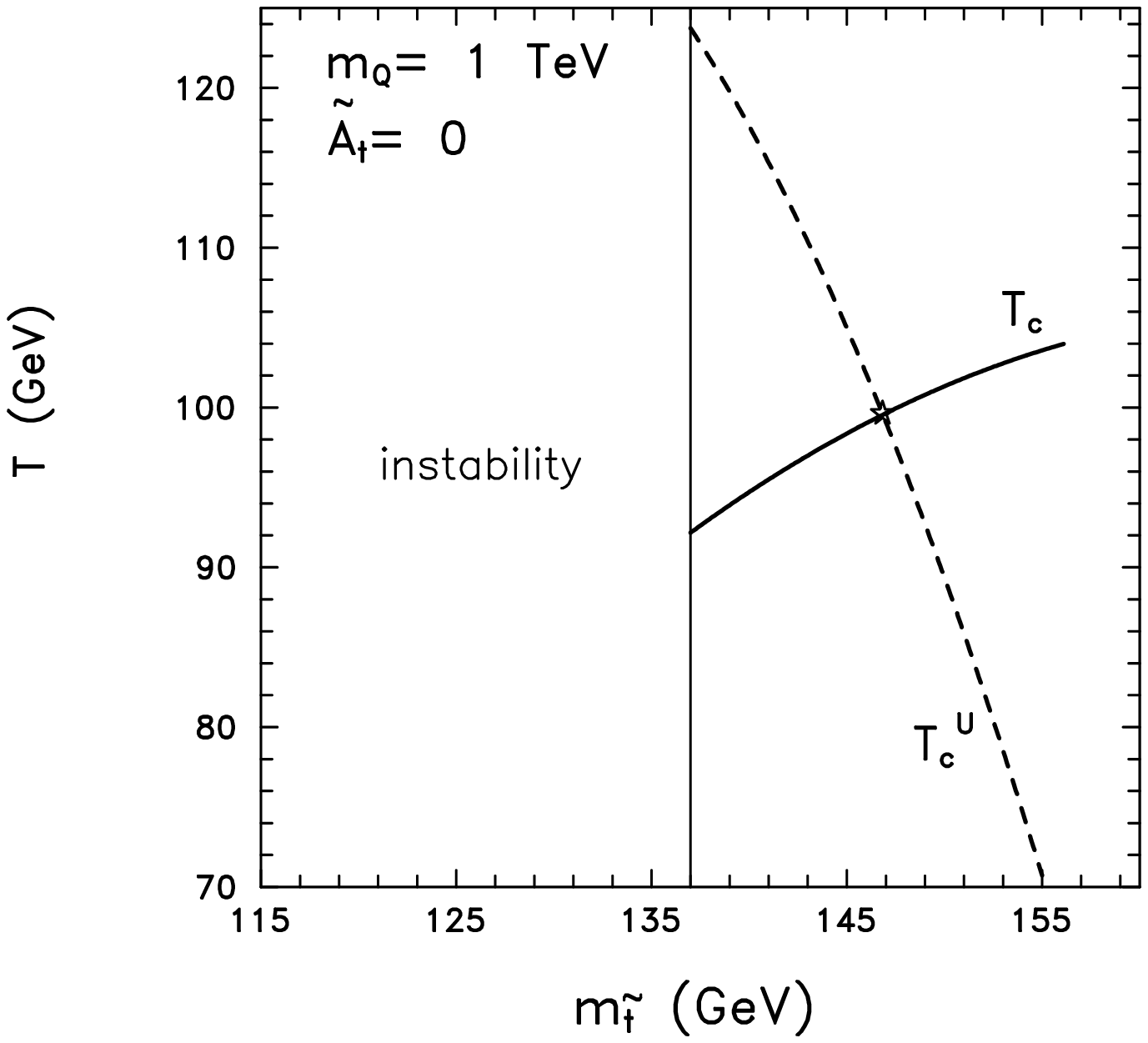,height=5.6cm,bbllx=3.5cm,bblly=3.5cm,bburx=17.5cm,bbury=16.5cm}
\psfig{figure=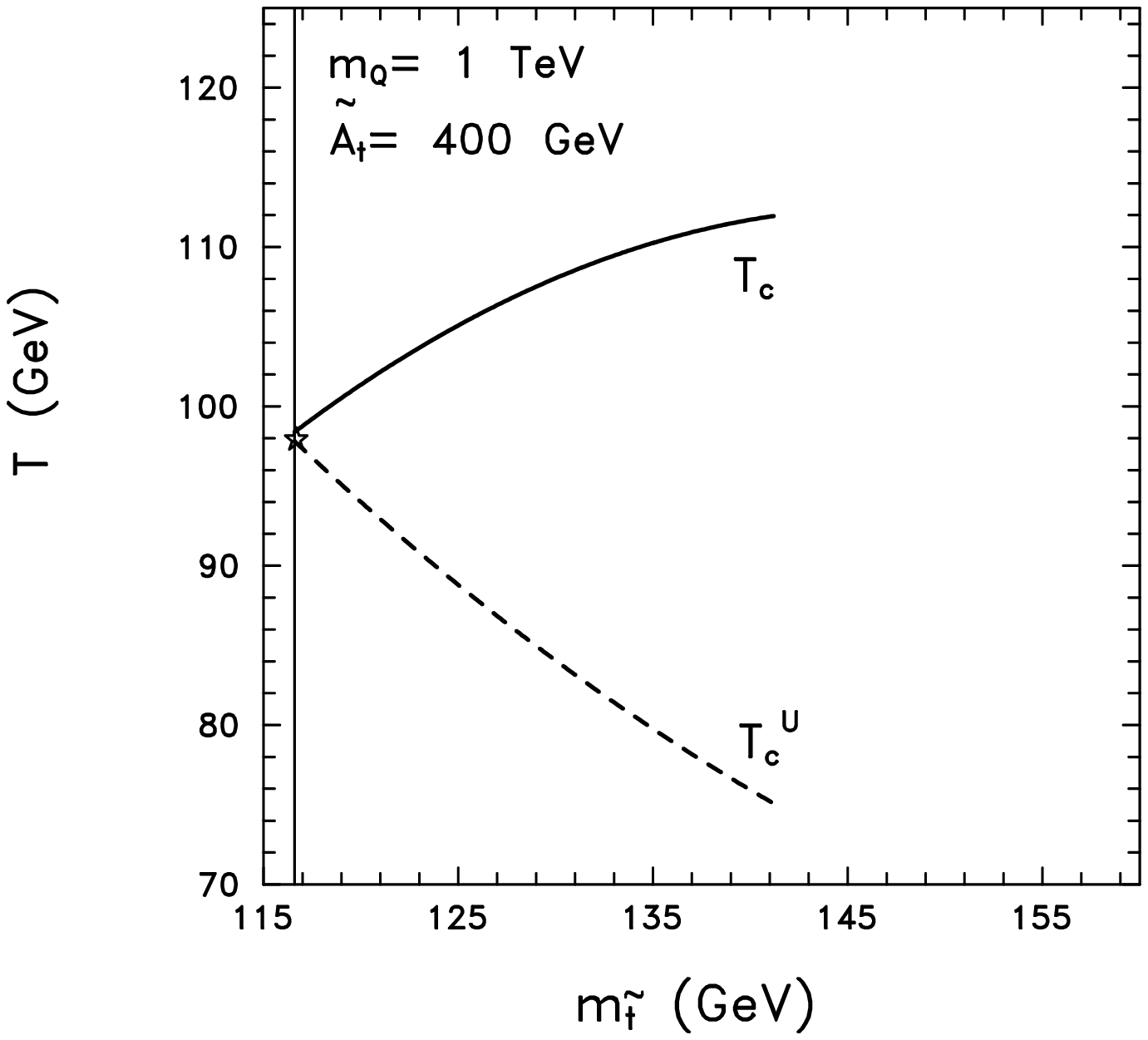,height=5.6cm,bbllx=3.5cm,bblly=3.5cm,bburx=17.5cm,bbury=16.5cm}}
\caption{Plot of the critical temperatures $T_c$ and $T_c^U$ for
$m_Q=1$ TeV, $\tan\beta=3$ and: $\widetilde{A}_t=0$, $m_H=89$
GeV, 
left panel; 
and,  $\widetilde{A}_t=400$ GeV, $m_H=94$
GeV, right panel.}
\label{fig2}
\end{figure}

For a given value of the supersymmetric parameters we have
computed $T_c$, the critical temperature along the $\phi$
direction, and $T_c^U$, the critical temperature along the $U$
direction. The comparison of them will provide information
about the cosmological evolution of the system. We have plotted
in Fig.~\ref{fig2} the critical temperatures as a function of
$m_{\widetilde{t}_R}$ for the values of the supersymmetric
parameters indicated in the caption, and for different values of
the mixing parameter $\widetilde{A}_t$. For a given point the
phase transition will happen first along the direction whose
critical temperature is higher. Then for $\widetilde{A}_t=0$ the
phase transition will proceed along the $\phi$ ($U$) direction
for $m_{\widetilde{t}_R}\simgt 147$ GeV
($m_{\widetilde{t}_R}\simlt 147$ GeV). For $\widetilde{A}_t=200$ GeV the
phase transition will proceed along the $\phi$ ($U$) direction
for $m_{\widetilde{t}_R}\simgt 137$ GeV
($m_{\widetilde{t}_R}\simlt 137$ GeV), and for
$\widetilde{A}_t=400$ GeV the
phase transition will proceed always along the $\phi$ ($U$) direction.
These results are in agreement with those of Fig.~\ref{fig1}.

Therefore, four different situations can arise:
\begin{itemize}
\item {\bf a)}
$T_c^U<T_c$ and $\widetilde{m}_U<m_U^{c}\equiv 
\left(m_H^2 v^2 g_s^2/12\right)^{1/4}$~\cite{CQW}. In this case the phase transition proceeds first along the $\phi$ direction and the field remains at
the electroweak minimum forever. This region is called {\em stability} region.
\item {\bf b)}
$T_c^U<T_c$ and $\widetilde{m}_U>m_U^c$. In this region the electroweak 
minimum is metastable ({\em metastability} region). It can be physically 
acceptable provided that its lifetime is larger than the age of the universe
at this temperature: $\Gamma_{\phi\rightarrow U}<H$~\cite{mqs2}.
\item {\bf c)}
$T_c^U>T_c$ and $\widetilde{m}_U<m_U^c$. In this case the $U$-phase transition
happens first and therefore it is physically acceptable provided that the
lifetime is shorter than the age of the universe: 
$\Gamma_{U\rightarrow \phi}>H$~\cite{moore}. This region is called {\em two
step} region.
\item {\bf d)}
$T_c^U>T_c$ and $\widetilde{m}_U>m_U^c$. This is the region of 
{\em instability} of the electroweak minimum. It is absolutely excluded.
\end{itemize}
\begin{figure}[htb]
\centerline{
\psfig{figure=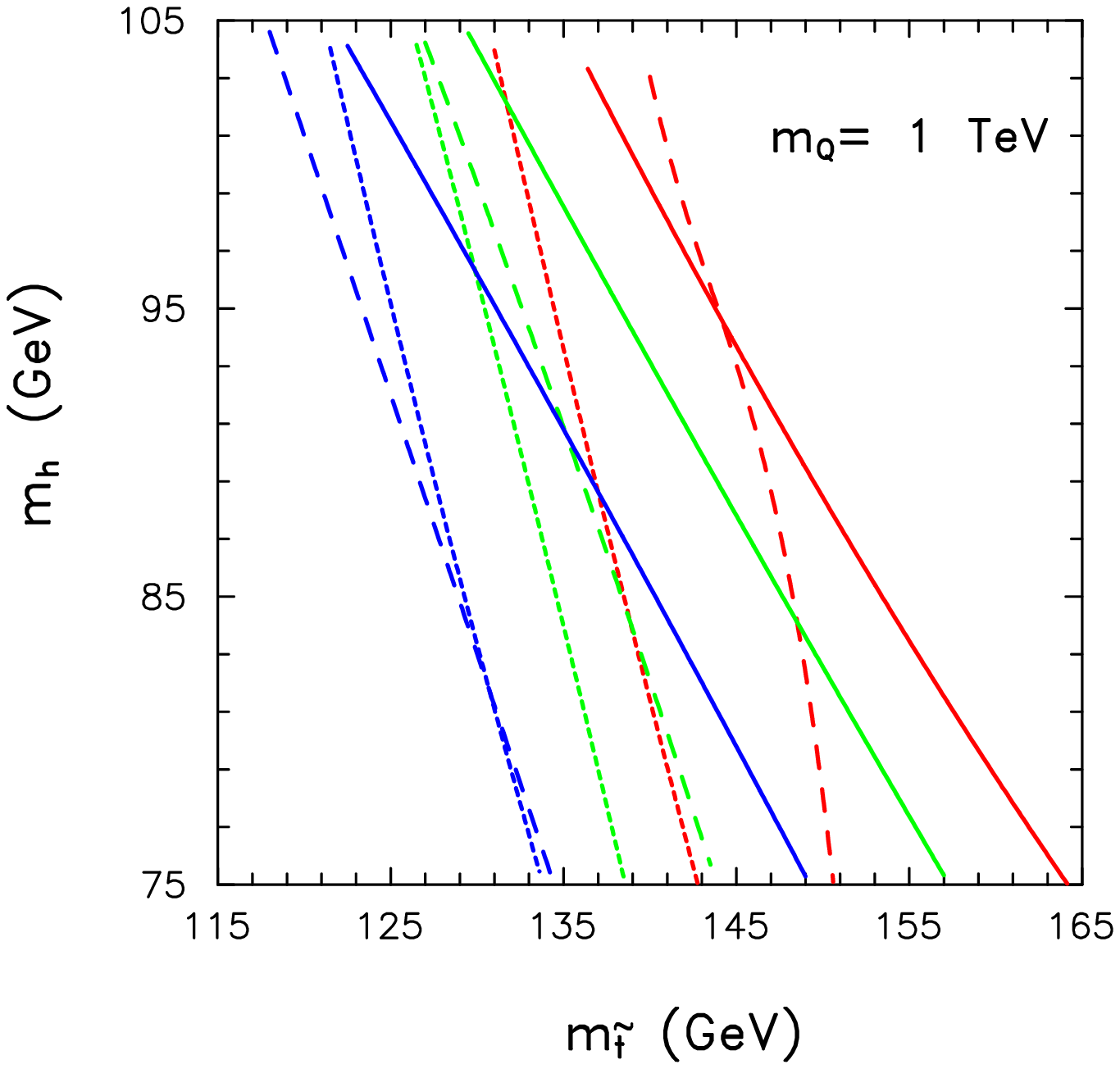,height=6.2cm,bbllx=5.9cm,bblly=5.cm,bburx=19.9cm,bbury=18.cm}
\psfig{figure=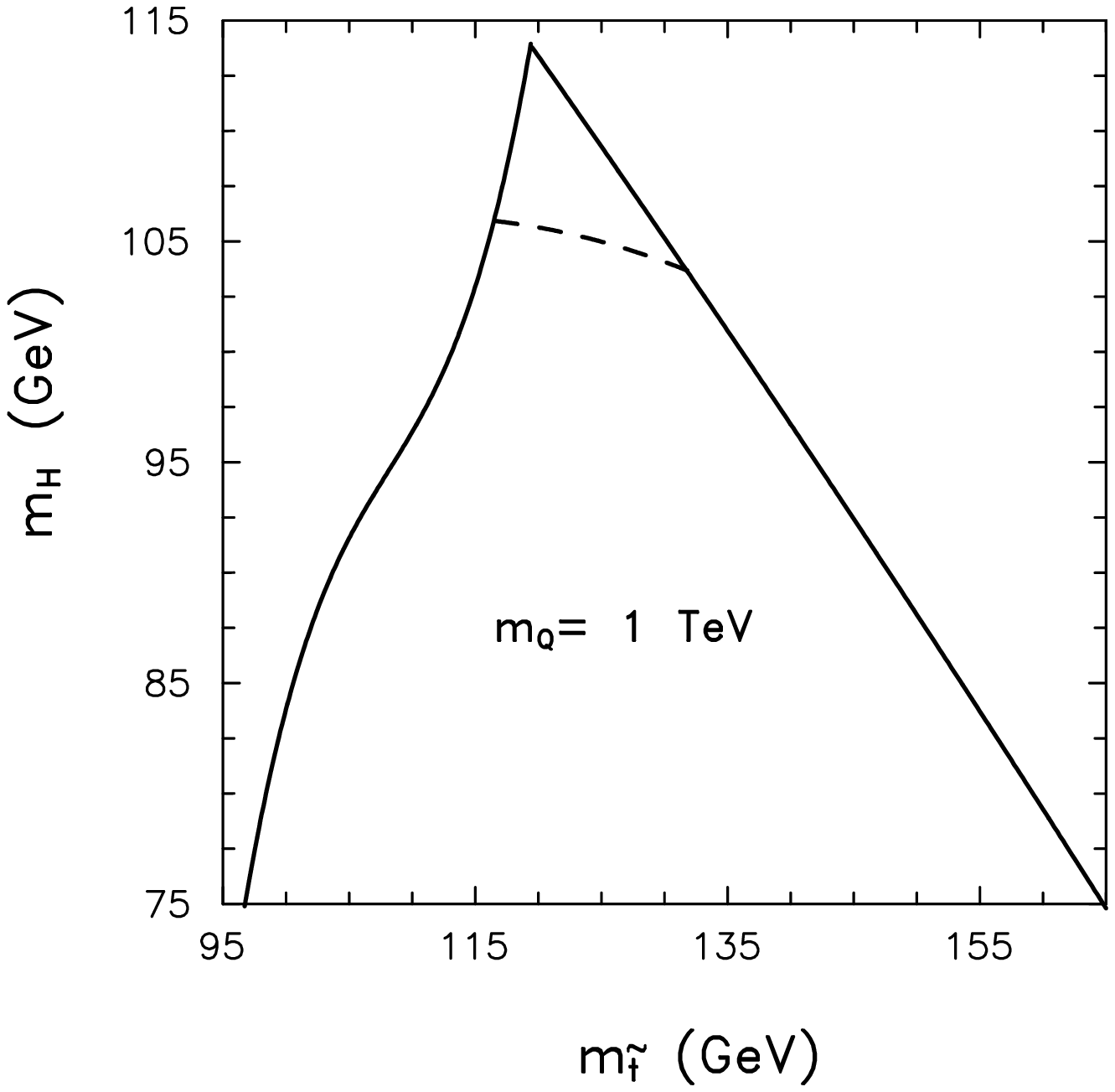,height=6.2cm,bbllx=5.cm,bblly=3.8cm,bburx=19.cm,bbury=16.8cm}}
\caption{Left panel: values of $m_h$, $m_{\st}$ for which $v(T_c)/T_c = 1$ 
(solid line), $T_c^U = T_c$ (dashed line), $\widetilde{m}_U = 
\widetilde{m}_U^c$ (short-dashed line), for $m_Q = 1$ TeV
and $\widetilde{A}_t=0,200,300$ GeV.
The region on the left of the solid line is consistent with a strongly
first order phase transition. A two step phase transition may occur in
the regions on the left of the
dashed line, while on the left of the short-dashed line, the physical
vacuum at $T = 0$
becomes metastable. The region on the left of both the dashed
and short-dashed lines leads to a stable color breaking vacuum state
at zero temperature and is hence physically unacceptable. Right panel:
The absolute region of stability in the ($m_H,m_{\widetilde{t}_R}$) plane, for
$m_Q=1$ TeV (below the dashed line) and for higher values of $m_Q$,
$\sim$ a few TeV (the region
inside the solid lines).}
\label{fig3}
\end{figure}
%
Fig.~\ref{fig3}, left panel, shows the region of parameter space 
consistent with a sufficiently strong phase transition for $m_Q=1$ TeV, 
$\widetilde{A}_t=0,200,300$ GeV. For low values of the
mixing, $\widetilde{A}_t \simlt 200$ GeV, 
case a) or c) may occur but, contrary to what
happens at one-loop, case b) is not realized. 
For the case of no mixing, this result
is in agreement with the analysis of~\cite{Schmidt}.
The region of absolute stability of the physical
vacuum for $\widetilde{A}_t \simeq 0$
is bounded to values of the Higgs mass of order
95 GeV. There is a small region at the right of the
solid line,
in which a two-step phase transition may
take place, for values of the parameters which
would lead to $v/T < 1$ for $T = T_c$, but may
evolve to larger values at some $T < T_c$ at which the 
second of the two
step phase transition into the physical vacuum takes place.
This region disappears
for larger values of the stop mixing mass parameter. 
For values of the mixing parameter $\widetilde{A}_t$ between
200 GeV and 300 GeV, both situations, cases b) and c) may
occur, depending on the value of $\tan\beta$.  
For large values of the stop mixing,
$\widetilde{A}_t > 300$ GeV,
a two-step phase transition does not take place. 
Fig.~\ref{fig3}, right panel, shows the absolute region of stability
for $m_Q\leq1$ TeV, below the dashed line, where we can see an absolute
upper bound on the Higgs mass $\sim$105 GeV. If we relax the condition 
on $m_Q$ and allow for values $\sim$2-3 TeV, then we get the bound, 
corresponding
to the region inside the solid lines, $\sim$115 GeV~\cite{cline}.

\section{Electroweak baryogenesis: two-loop enhancement}

The $CP$ violating current $\langle J_{CP}\rangle$ has been computed using 
triangle diagrams in a `Higgs insertion expansion'~\cite{CQRVW,Toni2}. Two 
kinds of diagrams are considered: stop mediated and Higgsino-gaugino 
mediated ones.
We find the former negligible due to the fact that we are considering 
left-handed stops much heavier than the temperature and therefore decoupled
from the thermal bath. The latter are found to be dominant and the amount of
generated baryon asymmetry is found, after taking for simplicity the thick wall
limit, $t_\omega=L_\omega/v_\omega\gg \tau_{\widetilde{t},\widetilde{H},...}
=1/\Gamma_{\widetilde{t},\widetilde{H},...}$, where $v_\omega$ is the wall velocity, as:
\begin{equation}
\frac{n_B}{s}\simeq f(m_Q,m_{\widetilde{t}},m_{\widetilde{H}},
\Gamma_{\widetilde{t}},\Gamma_{\widetilde{H}},\dots)\left\langle\frac{v^2(T_c)}
{T_c^2}\right\rangle\Delta\beta(T_c)
\label{nbs}
\end{equation}
where $\langle v^2(T)\rangle$ is the integral of the Higgs profile along its
radial coordinate, $\Delta\beta(T)$ the variation of the angle $\beta(T)$
along the bubble wall and the function $f$ comes from the integral of the
Feynman diagrams and integration of the diffusion equations. In this way the
main dependence on the phase transition is concentrated on the parameter
$\Delta\beta(T_c)$.

\begin{figure}[htb]
\centerline{
\psfig{figure=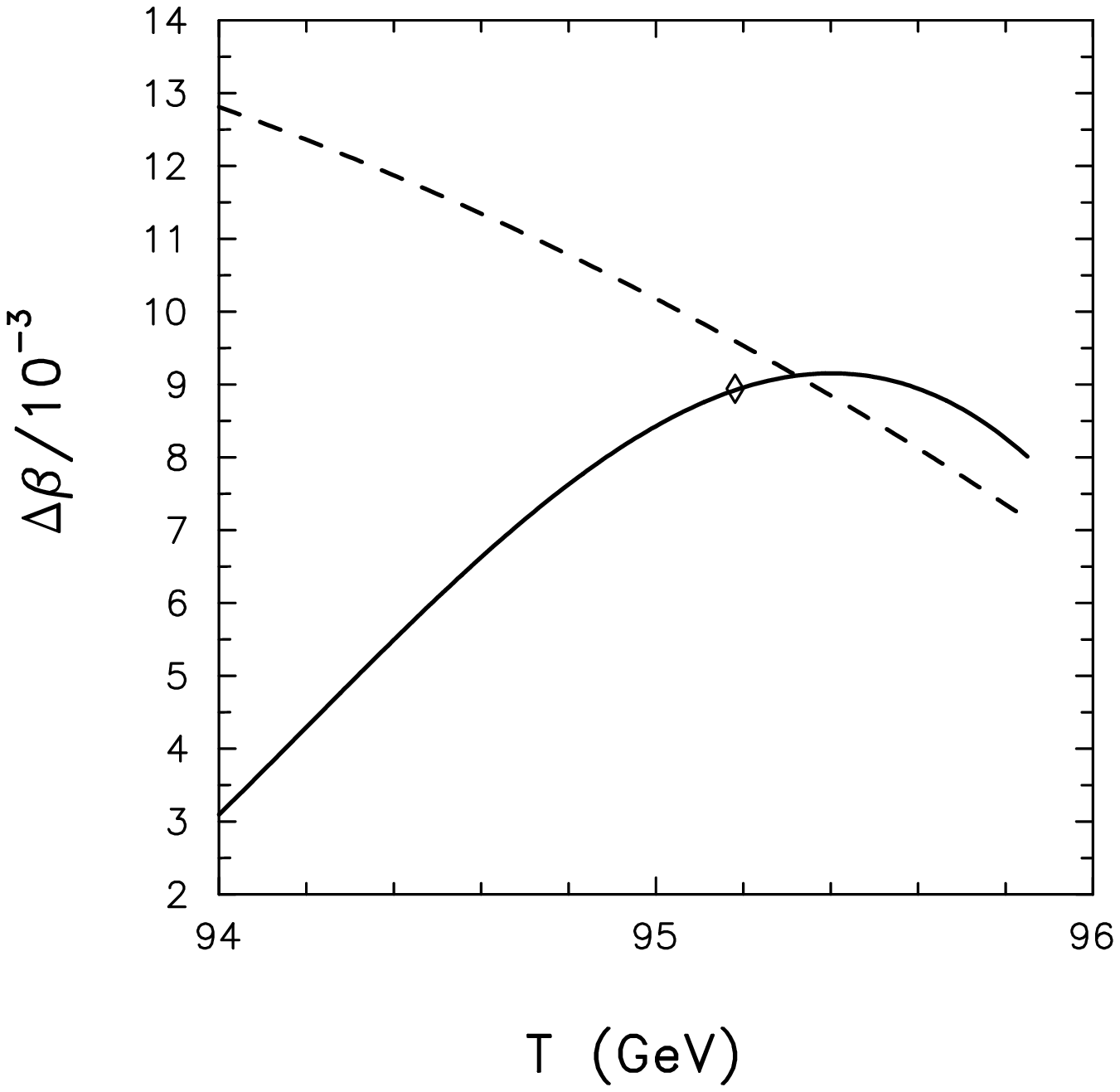,height=5.6cm,bbllx=4.5cm,bblly=3.5cm,bburx=18.5cm,bbury=16.5cm}
\psfig{figure=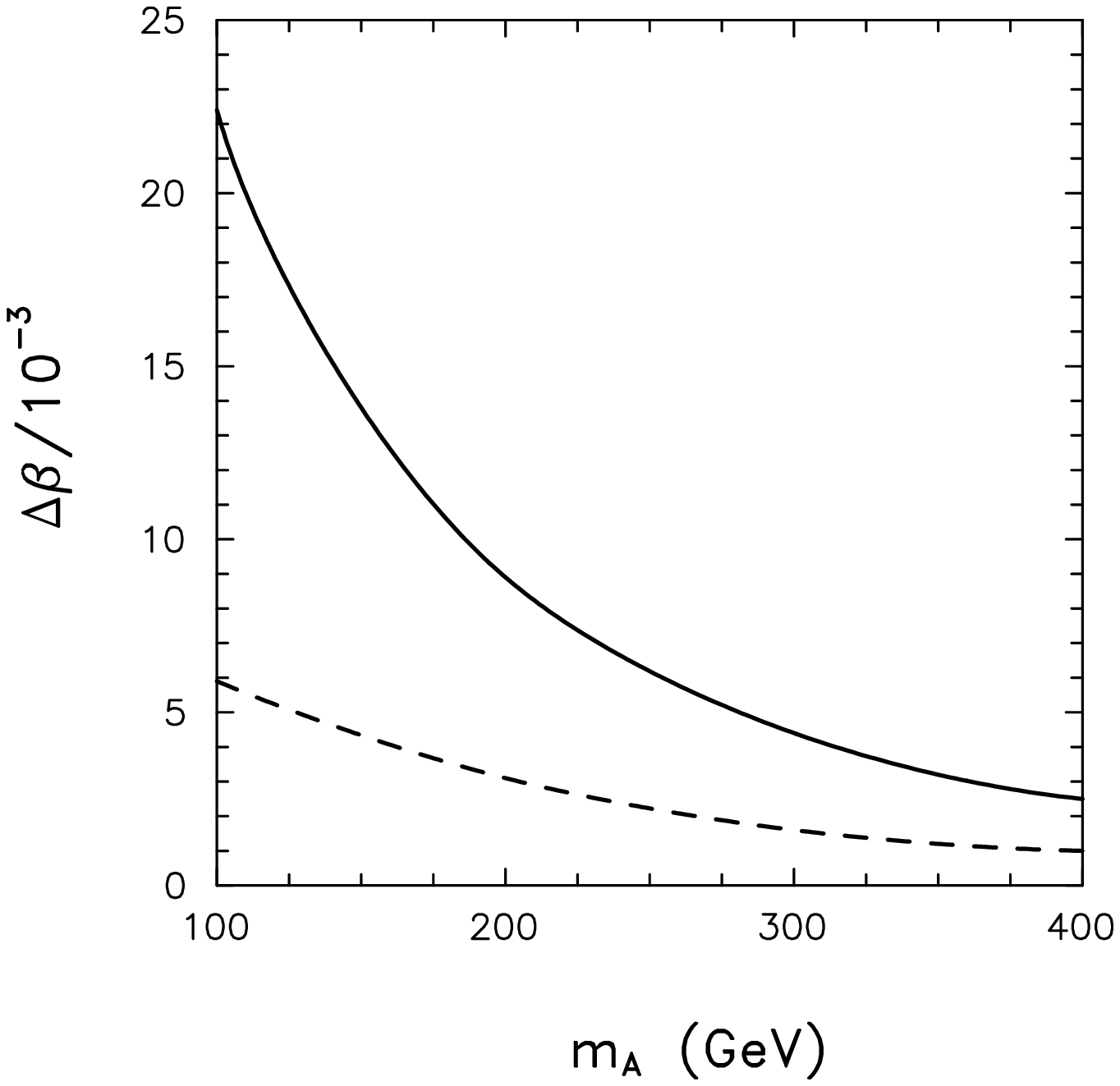,height=5.6cm,bbllx=4.5cm,bblly=3.5cm,bburx=18.5cm,bbury=16.5cm}}
\caption{Left panel: the parameter $\Delta\beta$ as a function of the 
temperature, for
$m_Q=1$ TeV, $\tan\beta=2.5$, $\widetilde{A}_t=0$, $m_{\widetilde{t}}=150$
GeV and $m_A=200$ GeV, from our numerical calculations (solid curve) and as 
obtained by pure potential energy considerations (dashed curve). The
diamond indicates the value of the critical temperature for the considered
case. Right panel: the parameter $\Delta\beta$ in the two-loop (solid curve)
and one-loop (dashed curve) approximations for the same values of the
supersymmetric parameters.}
\label{fig4}
\end{figure}
%
We have computed the bubble solutions of the MSSM~\cite{mqs} using, for the
sake of comparison, the one- and two-loop effective potential. We have 
confirmed the goodness of the thick wall approximation ($L_\omega T_c
\gg 10\, v_\omega$) and found, on $n_B/s$, a two-loop enhancement, with respect
to the one-loop result, which goes between one and two orders of magnitude.

\begin{figure}[htb]
\centerline{
\psfig{figure=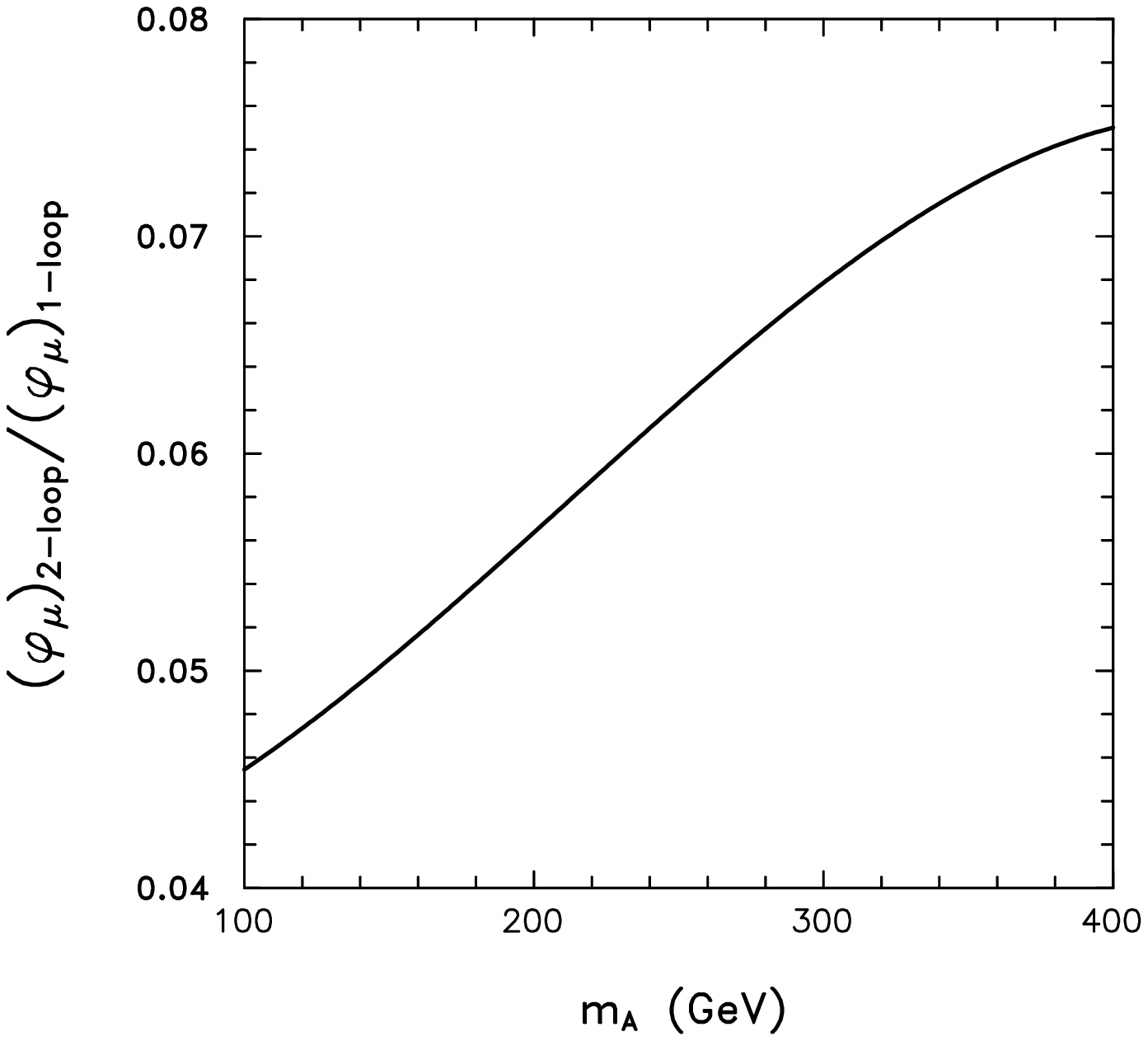,height=5.8cm,bbllx=5.5cm,bblly=4.5cm,bburx=19.5cm,bbury=17.5cm}
\psfig{figure=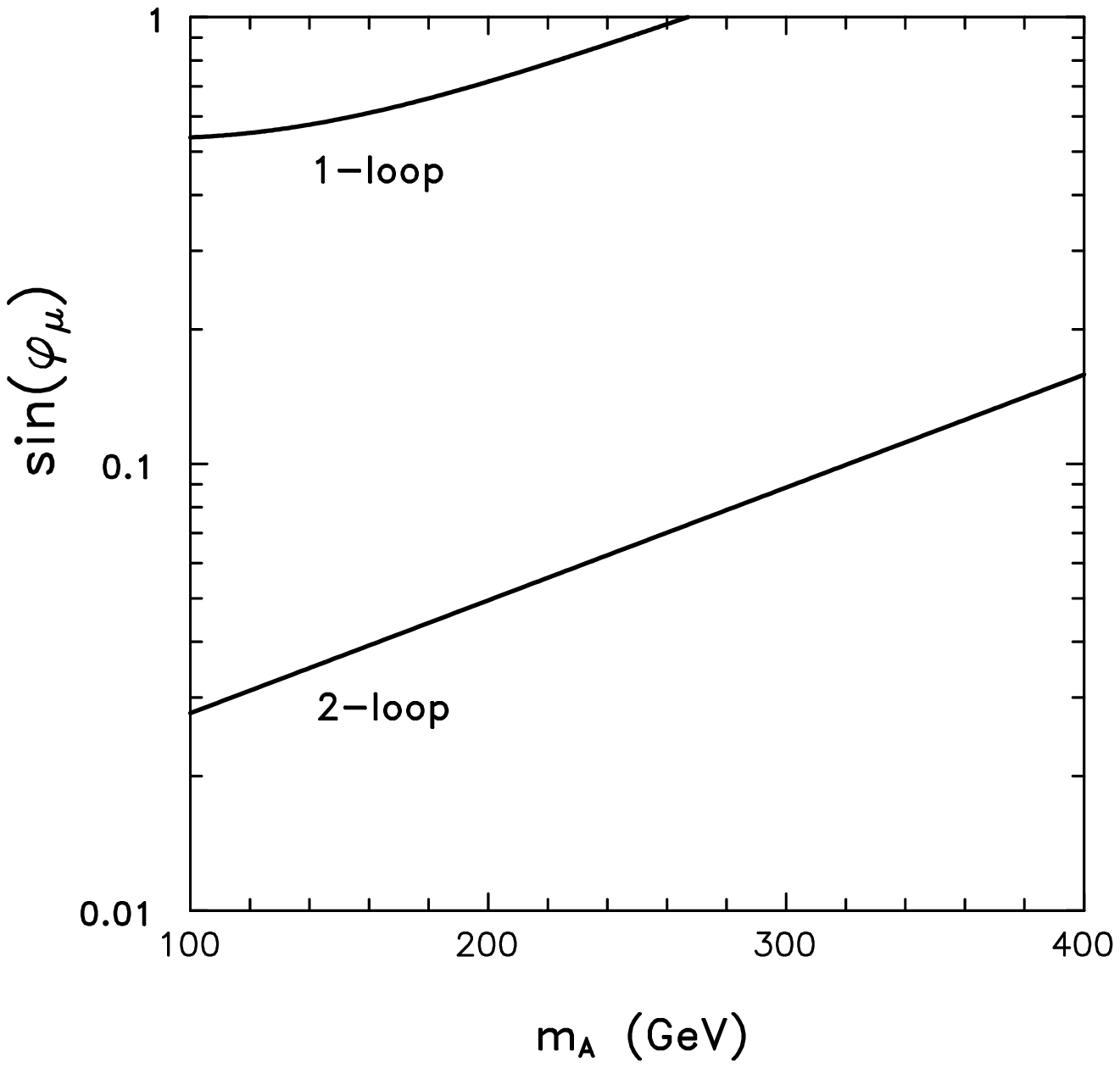,height=5.8cm,bbllx=5.5cm,bblly=4.5cm,bburx=19.5cm,bbury=17.5cm}}
\caption{Left panel: the two-to-one loop ratio (two-loop enhancement)
of $\varphi_\mu$ for the same values of supersymmetric parameters as in
Fig.~\ref{fig4}. Right panel: two- and one-loop calculations of the CP
violating parameter $\varphi_\mu$ as obtained by fixing $n_B/s$ to
its experimental value and using the calculation of (\ref{nbs}).}
\label{fig5}  
\end{figure}

In Fig.~\ref{fig4}, left panel, we plot $\Delta\beta(T)$ for the values of
the supersymmetric parameters indicated in the figure caption. The solid line
is the result of our numerical calculation and the diamond corresponds to the
actual value of the critical temperature for the considered case. For the sake 
of comparison we also plot (dashed line) the result assuming that the bubble
solution proceeds along the path of minimal potential energy. We can see that
there can be a large deviation between both results, the responsibility being
the contribution of the kinetic energy to the total action. In Fig.~\ref{fig4},
right panel, we plot the parameter $\Delta\beta(T_c)$ as a function of $m_A$
in the two-loop (solid line) and one-loop (dashed line) approximations. 
Notice that the value of $T_c$, as well as $\Delta\beta$, is different from
point to point, for different values of $m_A$. We can see that the ratio
$\Delta\beta_{2L}/\Delta\beta_{1L}$ varies from $\sim$ 3.6, for $m_A=100$ GeV,
to $\sim$ 2.5, for $m_A=400$ GeV. This enhancement, along with the two-loop
enhancement that we obtained for the quantity $v^2/T^2$, that can be quantified
from Fig.~\ref{fig1} as $\sim$ 6.7 for $m_A=100$ and $\sim$ 5.2 for $m_A=400$
GeV, provides a total enhancement in $n_B/s$, see Eq.~(\ref{nbs}), which can go
from $\sim$ 24, for $m_A=100$ GeV, to $\sim$ 13, for $m_A=400$ GeV.

Since the function $f$ in Eq.~\ref{nbs} goes linear in $\varphi_\mu$ (the
phase of the complex $\mu$), this enhancement translates into a smaller value
for $\varphi_\mu$ if we fix $n_B/s$ to its experimental value. Therefore the
value of $\varphi_\mu$ in the two-loop approximation is smaller than the
corresponding one in the one-loop approximation, for a fixed value of $n_B/s$.
This is shown in Fig.~\ref{fig5}, left panel, which is based on our numerical
solutions~\cite{mqs} and shows a two-loop enhancement $\sim$ 22 for $m_A=100$
GeV, and $\sim$ 13 for $m_A=400$ GeV, in agreement with our previous rough
estimates. The corresponding prediction for $\varphi_\mu$ from
(\ref{nbs})~\cite{CQRVW} is shown in Fig.~\ref{fig5}, right panel where we can
see that the CP violating parameter can be much smaller than that predicted by
the one-loop approximation.

\section{Conclusions}
Our results seem to point toward electroweak baryogenesis in the MSSM.
There is a baryogenesis window in the ($m_H$, $m_{\widetilde{t}}$) plane
summarized as:

\begin{itemize}
\item
The Higgs mass is constrained to $m_H\simlt 105$ GeV for $m_Q\simlt 1$ TeV.
This constrain can be relaxed if we allow left-handed stop masses to go to the
few TeV range, in which case the bound is $m_H\simlt 115$ GeV. We can compare
these numbers with the experimental bounds on the Higgs mass at LEP.
\begin{itemize}
\item
The present LEP bound is $\sim$95 GeV.
\item
At the end of the present run at LEP the bound on $m_H$ will increase to
$\sim$100 GeV.
\item
The LEP upgrade at $\sqrt{s}=200$ GeV, in 1999-2000, will explore the Higgs
mass up to $\sim$105-110 GeV.
\end{itemize}
As a consequence LEP will cover most of the Higgs window allowed by 
baryogenesis. 
\item
The stop mass is bounded to lie in the range $100\,{\rm GeV}\simlt 
m_{\widetilde{t}} \simlt m_t$. The lower bound can be relaxed a bit (to
80 GeV) if metastability solutions are satisfactory~\cite{mqs2}.
A wide region of stop masses will be tested at the next Tevatron run 
(starting end 1999). The whole region will be tested at TeV33.
\end{itemize}

To conclude, baryogenesis at the MSSM seems to be viable and provides a very
specific window in the space of supersymmetric parameters. This scenario
will be tested in the near future by collider experiments.

\section*{Acknowledgments}
We warmly thank our collaborators M.~Carena, J.~Moreno, A.~Riotto,
I.~Vilja and C.~Wagner, which whom recent works on the subject have been 
done. This work is partly supported by CICYT (Spain) under contract 
AEN98-0816.

\end{document}